Competitive Binding by Transcription Factors: A New Mechanism for Mendelian Dominance in Regulatory Genetic Pathways

Running title: Competitive Binding of Transcription Factors Causes Dominance

Adam H. Porter[*,†,1], Norman A. Johnson[*,†] and Alexander Y. Tulchinsky[†,‡]

[*]Department of Biology, University of Massachusetts-Amherst, Amherst, Massachusetts

[†]Graduate Program in Organismic and Evolutionary Biology, University of Massachusetts-Amherst, Amherst, Massachusetts

[‡]Department of Biology, SUNY New Paltz, New Paltz, New York

[1]Corresponding author: Department of Biology, University of Massachusetts-Amherst, Amherst, Massachusetts. Email: aporter@bio.umass.edu

**ABSTRACT** We report a new mechanism for allelic dominance in regulatory genetic interactions that we call binding dominance. We investigated a biophysical model of gene regulation, where the fractional occupancy of a transcription factor (TF) on the *cis*-regulated promoter site it binds to is determined by binding energy ($-\Delta G$) and TF dosage. Transcription and gene expression proceed when the TF is bound to the promoter. In diploids, individuals may be heterozygous at the *cis*-site, at the TF's coding region, or at the TF's own promoter, which determines allele-specific dosage. We find that when the TF's coding region is heterozygous, TF alleles compete for occupancy at the *cis* sites and the tighter-binding TF is dominant in proportion to the difference in binding strength. When the TF's own promoter is heterozygous, the TF produced at the higher dosage is also dominant. *Cis*-site heterozygotes have additive expression and therefore codominant phenotypes. Binding dominance propagates to affect the expression of downstream loci and it is sensitive in both magnitude and direction to genetic background, but its detectability often attenuates. While binding dominance is inevitable at the molecular level, it is difficult to detect in the phenotype under some biophysical conditions, more so when TF dosage is high and allele-specific binding affinities are similar. A body of empirical research on the biophysics of TF binding demonstrates the plausibility of this mechanism of dominance, but studies of gene expression under competitive binding in heterozygotes in a diversity of genetic backgrounds are needed.

# Introduction

Mendel (1866) coined the terms dominant and recessive to describe variants that respectively appear in 3::1 ratios in first-generation hybrid crosses. Wright (1934) proposed a plausible mechanism, demonstrating theoretically that dominance can arise as a natural consequence of functional allelic differences among enzymes that play roles in metabolic pathways. Alleles with reduced function tended to be recessive, and variation in the genetic background could modify the degree of dominance. Kacser and Burns (1981) cast Wright's mechanism into the language of enzyme kinetics and metabolic flux, a mechanism we will call *flux dominance*, and several studies have extended and modified it (e.g., Keightley and Kacser 1987; Keightley 1996; Bagheri and Wagner 2004). Since then several other mechanisms have been found to produce dominance, including negative regulatory feedback (Omholt *et al.* 2000), threshold-based reaction-diffusion systems (Gilchrist and Nijhout 2001), protein-protein interactions (Vietia *et al.* 2013) and epigenetic modifications (Li *et al.* 2012; Bond and Baulcombe 2014). In general, dominance arises because the relationship between the genotype and the phenotype it produces is non-linear (Gilchrist and Nijhout 2001; Vietia *et al.* 2013).

Empirical studies have shown that dominance is commonly found in loci involved in gene regulation. In particular, *trans*-acting alleles (e.g., transcription factors) commonly show dominance, whereas the *cis*-acting sites they regulate only rarely do (Li *et al.* 2012; Bond and Baulcombe 2014; Stupar and Springer 2006; Hughes *et al.* 2006; Wray 2007). The mechanism is unknown. We propose that this dominance is an inevitable consequence of differences in binding dynamics between *trans*-acting gene products as they compete for access to the *cis* sites they regulate. The degree of dominance thereby depends on differences in concentration and binding affinity of the *trans*-acting gene products. Such competitive binding interactions are integral to models of multifactorial gene regulation, including nucleosome–transcription-factor interactions (Tief *et al.* 2010, 2012), and repressor (Browning and Busby 2004) and microRNA function (Thomson and Dinger 2016), but have not been applied to allelic interactions. This form of dominance, which we term *binding dominance*, propagates through regulatory pathways and is modified by polymorphism at other loci in the pathway. Our findings apply to any *trans*-acting regulatory molecules interacting with *cis*-acting regulatory sites. Transcription-factor/promoter interactions meet these criteria well and we will develop the model using that language.

# Model

Biophysical models have long been used to study molecular interactions between DNA and molecules that bind to it (e.g., Gerland *et al.* 2002; Tulchinsky *et al.* 2014; Khatri and Goldstein 2015). The central premise of these models is that interactions between regulatory molecules and the sites they regulate behave according to the thermodynamic and kinetic principles that drive all molecular interactions. Consistent with empirical data (reviewed in Mueller *et al.* 2013), gene expression in these models only ensues while a transcription factor (TF) molecule is physically bound to the promoter of the regulated gene.

In our model, binding is a stochastic process determined by the free energy of association (–$\Delta G$) between a TF molecule and promoter, which we will call 'binding energy.' The fractional occupancy $\theta$ — the proportion of time a promoter is occupied by a TF molecule, and therefore the gene-expression level — depends on –$\Delta G$, and also on dosage $N_{TF}$, the number of free TF molecules available to bind when the promoter is unoccupied. We treat $N_{TF}$ and TF concentration as synonymous, using the nucleus as unit volume.

The biophysical model represents interacting TF molecules and the promoter sequence as strings of bits of arbitrary length, an approach based in statistical physics and information theory (Gerland *et al.* 2002). This method of abstraction permits characterization of molecular interactions at arbitrary scales, from the state space of electrostatic interaction among atoms to amino acid and nucleotide variation, and ultimately, to the genetic basis of variation in those molecules. The binding energy drops in steps of $-\Delta G_1$ as $m$, the proportion of mismatched bits over the length of the bitstring, increases (Tief *et al.* 2010). The haploid model, a parameter-reduced form of our model in Tulchinsky *et al.* (2014; see S1 Text), is

$$\theta = \frac{N_{TF}}{N_{TF} + \exp[-m\Delta G_1]} \tag{1}$$

We use the following notational conventions throughout. Interacting loci are labeled with letters A and B, with C included for 3-locus pathways. Subscripts indicate allelic variants as in Figure 1a; those before the letter (e.g., $_1$A, $N_{TF.1A}$) refer to promoter alleles and those after the letter (e.g., $A_1$) indicate gene-product (coding-region) alleles. Subscripts are dropped for homozygotes (e.g., AA; $N_{TF}$), and both subscripts are used when both sites vary for an allele (e.g., $_1A_1$, $_2A_2$, $N_{TF.1A1}$). Arrows indicate allele-specific regulatory interactions (e.g., $m_{A1 \rightarrow 1B}$ represents bitstring mismatches between TF allele $A_1$ and *cis*-site allele $_1$B).

## *Diploid model*

In diploids, the allelic forms $A_1$ and $A_2$ of the TF molecule (Figure 1a) compete for occupancy at both promoter sites $_1$B and $_2$B independently (Tulchinsky *et al.* 2014) and the total number of TF molecules is the sum of those from each TF allele copy ($N_{TF} = N_{TF.1A} + N_{TF.2A}$). Under TF competition, the fractional occupancy of $_1A_1$ on promoter site $_1$B in the presence of $_2A_2$ is

$$\theta_{1A1 \rightarrow 1B} = \frac{N_{TF.1A1}}{N_{TF.1A1} + \alpha_{2A2 \rightarrow 1B} \exp[-m_{1A1 \rightarrow 1B}\Delta G_1]} \tag{2a}$$

$$\alpha_{2A2 \rightarrow 1B} = 1 + N_{TF.2A2} \exp[m_{2A2 \rightarrow 1B}\Delta G_1] \tag{2b}$$

where $m_{1A1 \rightarrow 1B}$ and $m_{2A2 \rightarrow 1B}$ are the proportions of mismatches between the bit strings of $_1$B vs. $_1A_1$ and $_2A_2$ respectively, and $\alpha_{2A2 \rightarrow 1B}$ is the coefficient of competition with $_2A_2$ (Tulchinsky *et al.* 2014). Fractional occupancies of the other three interactions are calculated analogously.

The final expression level ($\phi$) is the sum of the fractional occupancies of the four TF-promoter pairs, scaled to the range [0, 1] such that there is no expression at minimal fractional occupancy,

$$\phi^* = \tfrac{1}{2}\left((\theta_{1A1 \rightarrow 1B} + \theta_{2A2 \rightarrow 1B}) + (\theta_{1A1 \rightarrow 2B} + \theta_{2A2 \rightarrow 2B})\right) \tag{3}$$

$$\phi = \max\left[\frac{\phi^* - \theta_{min}}{\theta_{max} - \theta_{min}}, 0\right] \quad (4)$$

where $\phi^*$ is the unscaled expression. Maximum fractional occupancy $\theta_{max} = N_{TF.sat}/(1+ N_{TF.sat})$ occurs when $m = 0$, and minimum fractional occupancy $\theta_{min}$ occurs when $m = 1$, for all TF-promoter pairs when both TF variants are at saturating concentration (i.e., $N_{TF.1A} + N_{TF.2A} \geq N_{TF.sat}$). As $\phi^*$ can be below $\theta_{min}$ when $N_{TF} < N_{TF.sat}$, we set the floor at $\phi = 0$. As a baseline for scaling purposes, we use dosages $N_{TF.1A} = N_{TF.2A} = N_{TF.sat}/2$ as the allele-specific saturating concentrations when $m = 0$ for both alleles.

### *Genotype-phenotype (G-P) map*

We treat the phenotype, $P$, as being proportional to the expression level of the *cis*-regulated locus, such that $P = k\phi$, and without loss of generality, treat that proportionality constant as $k = 1$, such that $P = \phi$.

In the biophysical model, the bit strings are abstract representations of information content that can characterize underlying genetic differences in the interacting molecules. Equations 1 and 4 therefore characterize the genotype-phenotype (G-P) map, the rules by which the phenotype is generated from the underlying genotype, as a function of binding energy and TF concentration.

### *Dominance*

Competition between TF alleles for binding to their *cis*-regulated sites creates conditions for allelic dominance (Tulchinsky *et al.* 2014). Following Wright (1934), we use $d = (P_{11} - P_{12})/(P_{11} - P_{22})$ as the dominance coefficient, where $P_{12}$ is the heterozygote phenotype and $P_{11}$ and $P_{22}$ are homozygote phenotypes; allele '1' of the respective locus is thereby the reference allele for which dominance is assessed. Allele 1 is codominant when $d = \frac{1}{2}$, completely dominant at $d = 0$ and completely recessive at $d = 1$.

If fractional occupancy cannot be measured separately for each allele, then $d$ must be assessed phenotypically. Even strong dominance becomes increasingly difficult to detect as $\phi$'s for homozygotes and heterozygotes of both alleles approach equality because the three genotypes will have very similar phenotypes; the trait will appear to be unaffected by these loci, or the degree of dominance will be obscured by sampling and measurement error. Detectability ($t$) is proportional to the absolute difference between the two homozygote phenotypes, such that $t = \kappa |P_{11} - P_{22}|$ with proportionality function $\kappa$. In a constant genetic background, $\kappa$ is some increasing function of the accuracy in the measurement of $P$ (or $\phi$) and the sample size of the study.

### *3-locus pathways: propagation and genetic background*

In a linear 3-locus pathway (Figure 1b), locus B codes for a second TF that binds to the promoter of locus C, such that there are two regulatory steps, A→B and B→C. The final phenotype is the fractional occupancy at locus C ($P = \phi_C$). The promoter and product sites of locus B together comprise a single allele (in this 3-locus, 2-allele model), and the doubly heterozygous B genotype is denoted $_1B_{12}B_2$. Competitive binding of the A alleles onto the two B alleles proceeds independently, creating two allele-specific fractional occupancy terms, $\phi_{1B1}$ and $\phi_{2B2}$, based on Equation 2. Expression of these B alleles yields separate $N_{TF.1B1}$ and $N_{TF.2B2}$ values, which we calculate as $N_{TF.1B1} = \phi_{1B1} N_{TF.B.sat}/2$ and $N_{TF.2B2} = \phi_{2B2} N_{TF.B.sat}/2$, such that maximal expression of the B locus yields $N_{TF.B.sat}$.

## Methods

We considered cases where fractional occupancy and therefore gene expression is maximal ($\phi = P = 1$) when binding is maximal ($m = 0$) and TF concentration is saturating, and that $\phi = P = 0.5$ when $m = 0.5$ at the same $N_{TF.sat}$. Analysis of the role of TF concentration requires scaling $-\Delta G_1$ to $N_{TF.sat}$ in order to meet these constraints. Substituting Equation 2 into Equation 3 and solving for $-\Delta G_1$, we used $-\Delta G_1 = 2 \ln[ N_{TF.sat} / (1 + N_{TF.sat}) ]$.

We report results from the cases where $N_{TF.sat}$ takes the values 10, 100 and 1000. To graphically illustrate the effects of detectability, we overlay the genotype-dominance maps with white opacity masks, grading from opaque at $t = 0$ through translucency to transparency at $t = 1$, where $t$ is the detectability parameter, with the effect of making the underlying genotype-dominance map increasingly visible as $t$ increases. As a heuristic, we treat scaling function $\kappa$ as a constant arbitrarily set to 4 with a maximum of $t = 1$; i.e., dominance is undetectable when homozygote phenotypes are equal and always detectable when their difference equals or exceeds 1/4.

In 3-locus pathways, we used Equations 2 and 4, with appropriate subscripts, to calculate $\phi_C$. For simplicity we assume $N_{TF.sat}$ is the same for both regulatory steps, i.e., $N_{TF.A.sat} = N_{TF.B.sat} = N_{TF.sat}$. All analyses were done using *Mathematica* (Wolfram Research, 2015).

## Results

We compare three types of polymorphism (Figure 1a). Polymorphism in the *cis*-regulated B locus is represented as AA→$_1$B$_2$B; that in the TF protein-coding region is A$_1$A$_2$→BB; and variation in TF dosage (i.e., allele-specific $N_{TF}$ as determined by upstream expression) is $_1$A$_2$A→BB. In the 3-locus AA→BB→CC pathway, we consider the propagation and detectability of dominance at locus A with respect to expression at downstream locus C ($\phi_C$) and explore genetic-background effects when loci B and C are polymorphic or have imperfect binding.

### *Genotype-phenotype maps*

The shapes of the G-P maps differ depending on which site is polymorphic. In the $_1$A$_2$A→BB case (Figure 2a-c) with maximal TF binding ($m = 0$), $\phi$ is low when both alleles are at low dosage ($N_{TF}$), climbing towards high expression as $N_{TF}$ of both alleles rises to saturating concentration $N_{TF.sat}$. The effect is very sensitive to $N_{TF.sat}$ such that the region of detectably lower $\phi$ is confined to the very bottom left corner of Figure 2c when $N_{TF.sat}$ is high. The drop-off in $\phi$ is proportional to their sum, $N_{TF}$, therefore perpendicular to the $N_{TF.1A} = N_{TF.2A}$ diagonal.

In the $A_1A_2 \rightarrow BB$ case (2d-f) at $N_{TF.sat}$, $\phi$ depends on competitive binding of the TF variants to the *cis*-sites they regulate (Equations 2 and 4). $\phi$ is high as long as either TF binds tightly ($m_{A1 \rightarrow B}$ or $m_{A2 \rightarrow B}$ is low), yielding a characteristic L-shaped ridge on the density plot, indicating dominance of the tighter-binding allele. Increasing $N_{TF.sat}$ (Figure 2e and 2f) broadens and flattens the ridge.

In the $AA \rightarrow {}_1B_2B$ case (Figure 2g-i), the expression of the two B-allele copies is additive (Equation 3) and at $N_{TF.sat}$, peak $\phi$ occurs when both alleles perfectly match the TF ($m_{A \rightarrow 1B} = m_{A \rightarrow 2B} = 0$). Expression falls away on both axes, leaving a characteristic arc on the density plot (Figure 2g), curving opposite the direction of the $A_1A_2 \rightarrow BB$ case. Increasing $N_{TF.sat}$ produces a more plateaued ridge that extends further out along the $m_{A \rightarrow 1B} = m_{A \rightarrow 2B}$ diagonal, visible as a more squared-off arc on the density plot (Figure 2h and 2i).

### *Dominance in expression level $\phi$*

Dominance at the A locus with respect to $\phi$ emerges in when variation occurs in the TF (the ${}_1A_2A \rightarrow BB$ and $A_1A_2 \rightarrow BB$ cases), with different patterns (Figure 3a-f). However, when variation occurs at the *cis* site (the $AA \rightarrow {}_1B_2B$ case) expression is always codominant ($d = 0.5$; not illustrated) due to the additivity of the products of locus B (Equation 3).

When TF binding varies (the $A_1A_2 \rightarrow BB$ case; Figure 3a-c), the TF allele with higher binding affinity (lower $m$) has a competitive advantage and dominant expression. The isoclines follow the diagonal when $m$ is low but flare at higher $m$ such that the competitive binding effect becomes much weaker. In this range the occupancy of each allele is so low that the TF's effectively cease to compete and the phenotype approaches additivity (i.e., diploid $\phi*$ of Equation 3 approaches haploid $\theta$ of Equation 1 as $m$ goes to 1). $N_{TF.sat}$ has a strong effect on dominance due to its effect on competition. When $N_{TF.sat}$ is high (Figure 3c), small changes in binding affinity can produce large changes in $d$, particularly when $m < 0.5$, whereas much larger changes in $m$ are required for the same effect at $N_{TF.sat} = 10$ (Figure 3a). Polymorphism in the B locus has no effect on the dominance of $A_1$ in the $A_1A_2 \rightarrow {}_1B_2B$ case.

When TF dosage varies (the ${}_1A_2A \rightarrow BB$ case; Figure 3d-f), the A allele with higher $N_{TF}$ is dominant. The isoclines spread linearly from the bottom left corner of the density plot, where $N_{TF}$ is low for both alleles, continuing into the region beyond the dotted line where total TF concentration is saturating ($N_{TF.1A} + N_{TF.2A} \geq N_{TF.sat}$). This dominance pattern is not substantially altered by $N_{TF.sat}$, nor is it by $m < 1$ provided that the TF coding region and the *cis*-site are homozygous. These plots are therefore not shown.

When TF dosage and binding affinity both vary (the ${}_1A_{12}A_2 \rightarrow BB$ case), the two sources of dominance interact cooperatively. Figure 3h shows the effect of allelic variation $N_{TF.1A1}$ and $N_{TF.2A2}$ under conditions where $m_{A1 \rightarrow B} = 0.1$, $m_{A2 \rightarrow B} = 0.2$, and $N_{TF.sat} = 100$. For orientation, Figure 3h represents the effects of varying dosage $N_{TF}$ for the binding-strength combination lying at the position of the circle in Figure 3b; the circles in the centers of Figure 3b and Figure 3h represent the same conditions. At this saturating concentration (i.e., $N_{TF.1A1} = N_{TF.2A2} = N_{TF.sat}/2$ at both circles), allele ${}_1A_1$ is dominant with $d = 0.291$. Along the x-axis in Figure 3g, increasing the dosage of the more tightly binding ${}_1A_1$ allele above $N_{TF.sat}/2$ increases its dominance, whereas decreasing its concentration pushes $d$ back towards codominance until ultimately dominance is reversed and ${}_1A_1$ becomes recessive. Along the y-axis, increasing the dosage of the $_2A_2$ allele also counteracts dominance of the $_1A_1$ allele, but the rate of change is much slower, and is only able to reverse the direction of dominance if $N_{TF.1A1}$ and $N_{TF.2A2}$ start well below $N_{TF.sat}/2$.

Dominance is more sensitive to binding affinity than to differences in dosage. Figure 3g reflects the same conditions as Figure 3e but with a 5-fold difference in allele-specific dosages, $N_{TF.1A1} = N_{TF.sat}/2$ and $N_{TF.2A2} = N_{TF.sat}/10$. For orientation, the orange crosses in Figures 3e and 3g share common parameter settings. Under these maximum-binding conditions, $_1A_1$ is dominant with $d = 0.17$. In Figure 3g, codominance is restored when binding of $_1A_1$ is reduced by ~20%, becoming recessive beyond that.

### *Detectability of dominance in the phenotype*

Figure 4a-f shows the dominance maps of Figure 3a-f overlaid by white opacity masks that obscure $d$ in proportion to the similarity of the expression levels in homozygotes. Existing dominance due to dosage differences in the $_1A_2A \rightarrow BB$ case is likely to be hard to detect unless $N_{TF.sat}$ is low and the dosages differ strongly (Figure 4d), and is likely to be detectable only in loss-of-expression alleles when $N_{TF.sat}$ is high (Figure 4e and 4f). Detectability is higher in the $A_1A_2 \rightarrow BB$ case especially when $N_{TF.sat}$ is low (Figure 4a). As $N_{TF.sat}$ increases (Figure 4b and 4c), the region of low detectability of dominance broadens in the high- and low-expression regions of the corresponding G-P maps (Figures 2e and 3f).

### *3-locus pathways*

Using a three-locus linear pathway (Figure 1b), we assessed the G-P maps and the dominance of the dosage ($_1A$) and binding ($A_1$) sites with respect to expression of locus C ($\phi_C$). We will call this dominance $d_{AC}$. We also examined the effects of genetic background by varying binding in the B$\rightarrow$C step.

The G-P map of the TF-dosage case (the $_1A_2A \rightarrow BB \rightarrow CC$ case) with $N_{TF.sat} = 10$ (Figure 5a) is a steeper version of the $_1A_2A \rightarrow BB$ map (Figure 2a), such that $\phi_C$ is nearly maximal unless $N_{TF.A}$ is very low for both A alleles. Higher values of $N_{TF.sat}$ yield such steep G-P maps at low dosage that only virtual double-knockout $_1A_2A$ genotypes are able to appreciably reduce locus C's expression (not illustrated). The G-P map for the TF-binding case (the $A_1A_2 \rightarrow BB \rightarrow CC$ case; Figure 5b shows $N_{TF.sat} = 10$) takes the same general form as the $A_1A_2 \rightarrow BB$ map (Figure 3e), but has a broad, high-expression plateau such that far greater A$\rightarrow$B mismatch is required for an equivalent reduction of $\phi_C$. At higher $N_{TF.sat}$ (not illustrated), the region of low expression becomes increasingly confined to the top right corner such that only very weak A$\rightarrow$B binding affects $\phi_C$ at maximal B$\rightarrow$C binding. The plateau becomes even broader and the shape squares off as it does for the $A_1A_2 \rightarrow BB$ maps of Figure 3d-f.

Dominance at the $_1A$ and $A_1$ sites propagates down the pathway to yield dominance with respect to $\phi_C$. In the $A_1A_2 \rightarrow BB \rightarrow CC$ case, the transition of $d_{AC}$ from dominant to recessive lies parallel to the $m_{A1 \rightarrow B} = m_{A2 \rightarrow B}$ diagonal when $m_{B \rightarrow C} = 0$ (Figures 5c and 6a), and increasing $N_{TF.sat}$ steepens the transition (Figures 5d and 6b). $d_{AC}$ is weaker and more sensitive to binding strength when $m_{B \rightarrow C} = 0.5$ but is slightly more detectable (Figure 6c and 6d). $d_{AC}$ drops rapidly between $0.5 \leq m_{B \rightarrow C} \leq 1$ and becomes very hard to detect, especially when $N_{TF.sat}$ is high (not shown). Here, without sensitive assays of expression, even unexpressed, completely recessive A alleles may go undetected.

Despite the differences in their G-P maps, dominance in the $_1A_2A \rightarrow BB \rightarrow CC$ case is almost identical to that of the $_1A_2A \rightarrow BB$ case seen in Figure 3d-f. However, its detectability (Figure 7a-f) is much weaker (e.g., compare Figure 7a to Figure 4d). It increases slightly when $m_{B \rightarrow C}$ = 0.5 (Figure 7c), but drops to become negligible beyond that (not shown). For higher levels of $N_{TF.sat}$, dominance will only be detectable when one of the A alleles is unexpressed (Figure 7b and 7d) unless assays are extremely sensitive.

Polymorphism in the genetic background can modify $d_{AC}$, but the magnitude of the effect depends on the background type. The effect is greatest in the $_1A_2A \rightarrow {_1B_2B} \rightarrow CC$ case, where dosage differences in TF locus A coexist with binding-site variation in the *cis* site of locus B. For illustration, we've chosen a combination where dosage of the $_1A$ allele is maximal ($N_{TF.1A}$ = $N_{TF.sat}/2$) and that of the 2A allele is low ($N_{TF.2A} = N_{TF.sat}/10$), at the position of the square in Figure 7a, such that dominance is strong and relatively easy to detect. Figure 6g shows the effect of binding variation in the A→B step, due to variation in the B-locus promoter ($m_{A \rightarrow 1B}$ vs. $m_{A \rightarrow 2B}$; the coding region of TF A is monomorphic), at this dosage combination. As overall A→B binding decreases ($m_{A \rightarrow B}$ increases), $d_{AC}$ increases (and becomes more detectable) until ultimately allele $_1A$'s becomes recessive. This effect is less pronounced as $N_{TF.sat}$ increases (Figure 6h), and also as the dosage differences decrease (not shown).

Other foreground/background combinations have weaker effects or none at all, and they mostly affect detectability. In Figure 6e and 6f, we show an example for the $A_1A_2 \rightarrow B_1B_2 \rightarrow CC$ case, where the genetic background consists of a high-functioning $B_1$ allele ($m_{B1 \rightarrow C}$ = 0) and a low-function $B_2$ allele ($m_{B2 \rightarrow C}$ = 0.9). Detectability is somewhat higher relative to the $A_1A_2 \rightarrow BB \rightarrow CC$ cases (Figure 6a and 6b, respectively) but the effect on $d_{AC}$ is negligible. In the $_1A_2A \rightarrow B_1B_2 \rightarrow CC$ case, detectability of $d_{AC}$ is largely determined by the dominant B allele in the B→C step, such that the genotype-dominance maps (not shown) are virtually indistinguishable from the $_1A_2A \rightarrow BB \rightarrow CC$ cases of Figure 7a and 7b. There is no effect on $d_{AC}$ of variation in the C-locus promoter (the $A_1A_2 \rightarrow BB \rightarrow {_1C_2C}$ and $_1A_2A \rightarrow BB \rightarrow {_1C_2C}$ cases; not illustrated), but it reduces detectability by reducing $\phi_C$.

## Discussion

We find that dominance emerges in regulatory genetic pathways due to competitive molecular interactions between transcription-factor variants in heterozygotes as they bind to their shared promoters. Alleles with higher competitive ability are inevitably dominant with respect to their contributions to fractional occupancy. Dominance effects extend to expression of downstream loci in multi-step pathways, and polymorphism therein can generate genetic background effects. However, this form of dominance is likely to be phenotypically detectable only when TF dosages or binding strengths are in the range where gene expression levels differ measurably among genotypes. We discuss each of these properties and their implications.

### *Binding dominance: a new mechanism for dominance*

Competition between transcription factors for binding to the promoter sites they regulate (Eq (2)); the $A_1A_2 \rightarrow BB$ and $_1A_2A \rightarrow BB$ interactions) represents a novel mechanism of dominance at the molecular level. The strength of the dominance depends on the biophysical properties of the interaction between TF molecules and the promoter sites to which they bind. When TF variants differ in their binding affinities ($-\Delta G$), the variant with higher affinity is dominant (Figure 3a-c). Dominance of the competing TF variants is also sensitive to TF availability ($N_{TF}$; Figure 3d-f). This is because when $N_{TF}$ is low, fractional occupancy is likewise low and there is little competition at the binding site; the allelic effects approach additivity. Conversely, at high $N_{TF}$, the more abundant TF allele more often occupies the promoter sequence, driving expression. In contrast, polymorphism at the downstream *cis*-regulatory site (AA→$_1$B$_2$B) cannot contribute to dominance. This is because expression of the *cis*-regulated gene product, or respectively the TF variant, proceeds independently for each allele and overall expression is their sum. In the 3-locus pathway, dominance in locus A can propagate down the pathway, such that A alleles can show dominance with respect to expression of locus C ($\phi_C$; Figures 5c, 5d, 6 and 7) as well as to locus B ($\phi_B$).

Binding dominance differs from the type of dominance that arises in metabolic pathways, which we call *flux dominance*, though the mechanisms of both are rooted in the biophysics of molecular interactions. In enzymes embedded in metabolic pathways, dominant alleles have higher rates of catalysis ($k_{cat}$), thus producing a higher flux from substrate to product, and the degree of dominance is proportional to the difference in $k_{cat}$ values (Kacser and Burns 1981; Keightley and Kacser 1987; Keightley 1996). Flux dominance is sensitive to substrate saturation of the enzyme (Bagheri-Chaichian *et al.* 2003), analogous to the way $N_{TF.sat}$ affects the degree of binding dominance through fractional occupancy. Flux dominance doesn't explain the effects of mutations at regulatory loci (Keightley 1996) because regulatory genetic pathways don't experience flux.

*Protein-assembly dominance* occurs when some subunits of complex proteins are expressed in inappropriate concentrations or have defective structures, disrupting the stoichiometry of protein assembly (Veitia 2003; Veitia *et al.* 2013). These represent downstream effects in the binding-dominance model, where subunit concentrations are determined by allele-specific $\phi_{1B1}$ and $\phi_{2B2}$, the expression levels of the B$_1$ and B$_2$ structural variants. The phenotype has a non-linear relationship to gene expression, or in our notation, $P = k\phi$ becomes $P = k(\phi_{1B1}, \phi_{2B2})$, where $k$ is now a function of the expression levels and binding properties of the other subunits in the complex.

*Feedback dominance* results from cases where a gene product autoregulates its expression. Omholt *et al.* (2000) analyzed feedback dominance using the biophysically relevant Hill (1910) equation that permits serially repeated promoter-site sequences; they considered only cases that lacked polymorphism in the TF coding region. Gene products could regulate either their own promoters (in our notation, $_1$A$_2$A→$_1$A$_2$A) or the promoters of an upstream TF ($_1$A$_2$A→$_1$B$_2$B→$_1$A$_2$A). These pathways resemble the $_1$A$_2$A→$_1$B$_2$B and $_1$A$_2$A→$_1$B$_2$B→CC cases for which we find dominance, suggesting that feedback dominance may ultimately prove to be a special case of binding dominance. To our knowledge, the effects of polymorphism in the coding regions, thus competitive binding, on feedback dominance remain unexplored.

*Diffusion dominance* arises in network-based regulation of ontogenetic diffusion gradients, including morphogen concentrations, their diffusion and decay rates, and the threshold concentrations necessary to initiate a phenotypic response (Gilchrist and Nijhout 2001). Allelic variation affecting any of these components can show dominance in network output. While we have presented our model in the context of TF-promoter interactions, its principles apply broadly to interactions between any genetically determined, interacting regulatory molecules. Our simple regulatory pathways represent elements in these more complex diffusion-based networks, and we expect that dominance due to competitive binding will be inherent in them.

### *Detectability and cryptic dominance*

Biophysical conditions that lead to especially high or low fractional occupancies, determining respectively the bottom left and top right corners of the G-P maps (Figuress 2, 5a and 5b), can mask dominance because the two homozygotes have very similar phenotypes. This can occur when $m$ is similar for both alleles, or when allele-specific dosage $N_{TF}$ is either high enough to saturate the binding site, or low enough that the binding site is rarely occupied by either allele. Even strong dominance at the level of molecular interactions can remain cryptic (e.g., compare Figure 3d-f to Figure 4d-f). When $N_{TF.sat}$ is high, only completely unexpressed $_1A$ or $_2A$ alleles will be detectable as recessive (Figure 4e and 4f) and moderate to strong dominance will likely go undetected. Likewise, when both TF alleles have similar binding affinities or dosages, the alleles will be nearly codominant, lying along the region of the diagonals of Figure 3a-f, but all individuals will also have nearly identical phenotypes. There, even polymorphism will be difficult to detect without genotyping; the degree of dominance may be of little practical importance in these cases anyway. Nevertheless, we predict that cryptic dominance will become apparent in assays of allele-specific expression levels (Mueller *et al.* 2013) in association with dosage and binding-strength variation.

Detectability of dominance in the 3-locus pathway (Figures 6 and 7) is lower than in the 2-locus pathway (Figure 4), because detectability is successively attenuated when it passes through $N_{TF}$ of downstream loci. In the 3-locus pathway, the A→B step determines $N_{TF.B}$. In general, $N_{TF}$ must be low for differences in $N_{TF}$ to affect expression (Figure 3a-c; this is also why low detectability is widespread in the $_1A_2A$→BB case of Figure 4d-f). It takes relatively large changes in expression in the A→B step to appreciatively change $N_{TF.B}$, and therefore to detect differences in expression at loci further downstream.

### *Effects of genetic background*

Polymorphism in the genetic background can enhance, obscure, or even reverse binding dominance. There are two types of background effects in the 2-locus regulatory interaction and several more in the 3-locus pathway. In the 2-locus pathway, dominance of coding-site ($A_1$, $A_2$) alleles at the TF locus is unaffected by polymorphism in the *cis*-regulated locus (i.e., $d_{A1A2 \to 1B2B} = d_{A1A2 \to BB}$). However, when allele-specific TF dosage and binding affinity ($N_{TF}$ and $m$) are permitted to vary in the $_1A_{12}A_2$→BB case, dominance of coding-site TF variants is affected by polymorphism in their promoters (Figure 3g) and vice versa (Figure 3h). For a given TF coding-region ($A_1A_2$) heterozygote, dominance modification is asymmetrical, being more effective when the dosage of the tighter binding A allele ($N_{TF}$) is varied (Figure 3h). In contrast, for a given dosage ($_1A_2A$) heterozygote, changes in binding affinities of either allele have effects of similar magnitude (Figure 3g).

In the 3-locus pathway, detectability of $d_{AC}$ is further modified by binding strength in the B→C step, such that it is least attenuated when $m_{B \to C} = 0.5$ (for $N_{TF.sat} = 10$, compare Figure 6a and 6c, also Figure 7a and 7c; for $N_{TF.sat} = 100$, compare Figure 6b and 6d, also Figure 7b and 7d).

This is where the G-P map for the B-locus TF coding region is steepest (Figure 2d-f), therefore where $|\phi_{C.11} - \phi_{C.22}|$ (the denominator of $d_{AC}$) is greatest. $d_{AC}$ becomes almost undetectable when $m_{B \to C}$ is high because G-P maps are nearly flat there (Figure 2d-f), such that the underlying 2-locus dominance is nearly undetectable (Figure 4b and 4c). Polymorphism at the coding site of locus B (the $A_1A_2 \to B_1B_2 \to CC$ and $_1A_2A \to B_1B_2 \to CC$ cases) modifies detectability only negligibly (Figures 6g, 6h, 7g and 7h), because expression at the B→C step incorporates dominance of the tighter-binding allele. Modifying of binding strength $m_{B \to C}$ by changing the C-locus promoter has the same effect on $d_{AC}$ as does changing the B-locus coding region, but without the effect of dominance in the BB→$_1C_2C$ case because expression there is additive.

Flux dominance is similarly sensitive to allelic substitutions that occur up to several steps removed along a metabolic pathway (Kacser and Burns 1981; Keightley 1996). Bagheri-Chaichian *et al.* (2003) show that the downstream dominance effects are sensitive to enzyme saturation at intermediate steps, much as we see in binding-site saturation in regulatory pathways (Figures 6 and 7). Feedback dominance likewise shows downstream effects (Omholt *et al.* 2000) in pathways with the structure $_1A_2A \to BB \to (_1A_2A$ & $CC)$, i.e., where the product of locus B co-regulates a downstream locus C as well as upstream locus A. In this case, dominance of the $A_1$ allele is detectable in the expression of locus C. Omholt *et al.* (2000) did not directly assess attenuation of the signal due to saturation at intermediate steps; rather, they noticed and excluded it by considering only cases where homozygotes showed differences >25%.

Binding dominance is likely to interact with flux dominance. When locus B codes for a metabolic enzyme, flux dominance of allele $B_1$ can be modified in $_1A_2A \to B_1B_2$ or $A_1A_2 \to B_1B_2$ interactions, provided that regulatory changes in B's expression levels affect enzyme saturation in the three B-locus genotypes. Polymorphism in both the promoter and product site of the B locus, i.e., the $_1A_2A \to _1B_{12}B_2$ and $A_1A_2 \to _1B_{12}B_2$ cases, should further influence $B_1$'s flux dominance by further changing relative allozyme concentrations. Conversely, we expect changes in allozyme concentration or $k_{cat}$ due to variation in $_2B$ or $B_2$ to modify, mask or expose binding dominance at $_1A$ or $A_1$ when $d_{A1}$ is assessed using genotype-specific fluxes in the metabolic pathway.

Beyond the regulatory pathway, transcription factors interact with other molecules in the cell that may be influenced by genetic background. These include direct interactions with proteins that regulate TF availability, spurious DNA, RNA or protein binding, and indirect effects of physiological conditions such as pH (Mueller *et al.* 2013). These affect the $N_{TF}/N_{TF.sat}$ ratio but have negligible effect on dominance and its detectability: the isoclines of Figure 3d-f and the detectability gradients of Figure 4d-f are linear, therefore constant with respect to this ratio. However, dominance may be modified in cases where TF variants differ in their responses to the non-specific background or are regulated differently (i.e., $A_1A_2$ cases with properties closer to the $_1A_{12}A_2$ case). For analytical convenience in this study, these secondary binding effects are subsumed into $N_{TF}$ (see parameter reduction in S1 Text). The unreduced model of Tulchinsky *et al.* (2014) may be necessary in the design and interpretation of experiments.

### *Empirical studies*

Consistent with the competitive binding model, *cis*-site heterozygotes typically show additive expression whereas *trans* heterozygotes commonly show dominance (Wray 2007; Guo *et al.*

2008; Tirosh *et al.* 2010; Zhang *et al.* 2011; Gruber *et al.* 2012; Meiklejohn *et al.* 2014), although some *cis*-site polymorphisms show patterns of dominance as well (Guo *et al.* 2008; Lemos *et al.* 2008). Our modeling suggests the possibility that unidentified polymorphism in regulatory loci upstream may be involved in at least some of these exceptions. Motifs with variable numbers of binding-site repeats in the promoter region could also potentially produce binding dominance and even overdominance, as they do in feedback dominance (Omholt *et al.* 2000).

Mueller *et al.* (2013) review empirical work on the biophysics of fractional occupancy in regulatory interactions. Gene expression is highly correlated with fractional occupancy of TFs on their binding sites, as our model assumes. Site-specific mutagenesis, using a variety of techniques for measuring binding affinity at primary vs. secondary (likely to be spurious background) binding sites, reveals strong differences in binding affinity among artificial promoter-region alleles ($_1B$ and $_2B$ alleles, in our notation). Some of these techniques are themselves based on measures of competitive binding among sites. Gaur *et al.* (2013) review studies demonstrating that TF and promoter-region alleles show significant patterns of allele-specific gene expression in diverse model organisms. Nevertheless, to our knowledge, allelic variation in TF binding affinity and concentration, in diverse genetic backgrounds, with respect to its effects on competitive binding and heterozygote gene expression remain to be studied.

### *Concluding remark*

In the discovery and documentation of regulatory architectures that drive gene expression, it has been necessary and appropriate to use inbred lines and careful breeding designs in model organisms to control for heterozygosity and to homogenize the genetic background. Outside of the laboratory, polymorphism is ubiquitous. Our understanding of gene regulation must account for it as we learn to predict and manipulate gene expression in the face of multilocus heterozygosity, and ultimately as we design and implement new regulatory architectures, in diverse systems of importance in medical, agricultural and fundamental research. A comprehensive, quantitative, mechanistically robust theory of Mendelian dominance will likely be required, and binding dominance is likely to be a significant component of it.

## Acknowledgments
We thank C. Babbitt for valuable comments on the manuscript.

**Figure legends**

**Figure 1** Diploid 2- and 3-locus regulatory pathways with competitive transcription factor (TF) binding. (a) 2-locus pathway: TF locus A codes for the TF protein that regulates the expressed locus B. Locus A can vary at its promoter (alleles $_1A$ & $_2A$), the coding region (alleles $A_1$ & $A_2$) or both; locus B varies only at the promoter ($_1B$ & $_2B$). The dosages of each of the TF alleles ($N_{TF.1A}$ and $N_{TF.2A}$) are determined by their promoter sequences. Subscripts are dropped for homozygotes. (b) 3-locus model: As in the 2-locus pathway, except that locus B codes for a second TF that goes on to regulate expression of locus C.

**Figure 2** Genotype-phenotype maps in the 2-locus regulatory pathway with competitive transcription factor (TF) binding. Genotype-phenotype maps, shown as density plots of expression ($\phi$), equivalent to phenotype in this model ($\phi = P$). Rows: Three saturating TF concentrations ($N_{TF.sat}$) at maximal binding ($m_{A1 \to B} = m_{A2 \to B} = 0$). (a-c) Effects of allelic variation in TF dosage, scaled to $N_{TF.sat}$. (d-f) Effects of allelic variation in the TF coding region expressed as mismatch ($m$) with a homozygous *cis*-site promoter BB. (g-i) Effects of allele-specific variation in the *cis*-site, holding the homozygous TF. Isoclines throughout represent intervals of 0.1 and the black isocline represents $\phi = 0.5$.

**Figure 3** Genotype-dominance maps, shown as density plots. Dominance ($d$) is with respect to the TF allele; $A_1$ (or $_1A$) is dominant in the blue region and recessive in the red. (a-c) $A_1A_2 \to BB$ cases: dominance at the TF coding region. $d$ as a function of the degree of mismatch ($m$) between a homozygous *cis* site and competing TF-coding alleles $A_1$ and $A_2$, for three saturating TF concentrations ($N_{TF.sat}$); the tighter-binding allele (low $m$) is dominant. The circle in panel 3b has the same $m$ and $N_{TF}$ values as the circle in panel 3h. (d-f) $_1A_2A \to BB$ cases: dominance in TF dosage. $d$ as a function of expression level of the two TF alleles $_1A$ and $_2A$, expressed as a fraction of the saturating TF dosage ($N_{TF.sat}$). TF dosage is saturating ($N_{TF.1A} + N_{TF.2A} \geq N_{TF.sat}$) above the dotted diagonal line. The higher-dosage TF is dominant. (g, h) $_1A_{12}A_2 \to BB$ cases. (g) Dominance as a function of the degree of mismatch in the heterozygous genetic background where TF alleles differ in dosage, for the case where $N_{TF.1A1} = N_{TF.sat}/2$ and $N_{TF.2A2} = N_{TF.sat}/10$. The orange crosses in this and panel 3e have the same $m$ and $N_{TF}$ values. Above the dotted line at $m_{A2 \to B} = 0.82552$, $\phi_{aa} = 0$ for the $_2A_2$ homozygote and $d = (\phi_{AA} - \phi_{Aa})/\phi_{AA}$, marking an discontinuity on the map. (h) Effect of allele-specific concentration in TF in a heterozygous genetic background, where the TF's coding region is heterozygous ($m_{1A1 \to B} = 0.1$, $m_{2A2 \to B} = 0.2$, $N_{TF.sat} = 100$). The circle has the same $m$ and $N_{TF}$ values as the circle in panel 3b. Isoclines throughout represent intervals of 0.1 and the thicker white isocline denotes $d = 0.5$.

**Figure 4** Detectability of dominance with respect to genotype. White opacity-gradient masks overlay the genotype-dominance maps of the corresponding panels of Figure 3a-f, such that the intensity of underlying color represents the detectability of dominance. Isoclines throughout represent intervals of 0.1 and the thicker white isocline denotes $d = 0.5$.

**Figure 5** Downstream effects of dominance in the 3-locus pathway: dominance of locus A with respect to expression at locus C. (a,b) Genotype-phenotype maps with $N_{TF.sat} = 10$. (c,d) Genotype-dominance maps of $A_1A_2 \to BB \to CC$, with $m_{B \to C} = 0$. (c) The dotted lines mark discontinuities: at high $m_{A1 \to B}$ beyond the blue dotted line, $\phi_{AA} = 0$ for the $A_1A_1$ homozygote such that $d = \phi_{Aa}/\phi_{aa}$; at high $m_{A2 \to B}$ above the red dotted line, $\phi_{aa} = 0$ for the $A_2A_2$ homozygote such that $d = (\phi_{AA}-\phi_{Aa})/\phi_{AA}$; beyond both lines in the top right corner, $d$ is undefined.  Isoclines throughout represent intervals of 0.1.

**Figure 6**  Genetic background effects on dominance in the 3-locus pathway.  Pattern and detectability of $d$ at the coding site of TF locus A in the 3-locus pathway with respect to expression at downstream locus C ($\phi_C$), when TF locus B varies in its coding site.  White opacity-gradient masks overlay the genotype-dominance maps, proportionally obscuring regions of low detectability.  Notation: $m_{B \to C} = \{x,y\}$ means that $m_{B1 \to C} = x$ and $m_{B2 \to C} = y$; locus B is homozygous when x=y. (a-f) Effects of varying binding in the B→C interaction by changing the coding site of the B-locus TF.  The dotted lines mark discontinuities: at high $m_{A1 \to B}$ beyond the blue dotted line, $\phi_{AA} = 0$ for the $A_1A_1$ homozygote such that $d = \phi_{Aa}/\phi_{aa}$; at high $m_{A2 \to B}$ above the red dotted line, $\phi_{aa} = 0$ for the $A_2A_2$ homozygote such that $d = (\phi_{AA}-\phi_{Aa})/\phi_{AA}$; beyond both lines in the top right corner, $d$ is undefined. (g, h) Effects of changing the B-locus promoter region in the $_1A_2A \to _1B_2B \to CC$ case, with detectability overlays.  Low binding in the A→B step reverses the dominance pattern due to dosage differences of the A alleles. (g) $m_{A \to 1B}$ vs. $m_{A \to 2B}$; $N_{TF.sat} = 10$, $N_{TF.1A} = N_{TF.sat}/2$, $N_{TF.2A} = N_{TF.sat}/10$.  The orange square at the origin represents the same parameter conditions as the orange square in Figure 7a; i.e., panel 6g represents a projection from Figure 7a taken at a point where the dosage of the $_1A$ allele is maximal and that of the $_2A$ allele is very low.  The dotted lines separate regions where the expression levels of either or both alleles of locus B ($\phi_{1B}$ and $\phi_{2B}$) equal 0 in the genotypes $_1A_2A$ and $_2A_2A$.  Those combinations are shown in Figure S1.  (h) Same, but with $N_{TF.sat} = 100$.  Isoclines throughout represent intervals of 0.1.

**Figure 7**  Pattern and detectability of dominance of allele-specific concentration differences of TF locus A in the 3-locus pathway, in relation to saturating TF concentration and binding strength in the B→C step.  (a-d) White opacity-gradient masks overlay the genotype-dominance maps, proportionally obscuring regions of low detectability.   Notation: $m_{B \to C} = \{x,y\}$ means that $m_{B1 \to C} = x$ and $m_{B2 \to C} = y$; locus B is homozygous when x = y. (a) The orange square represents the same parameter conditions as the orange square in Figure 6g, i.e., Figure 6g examines genetic-background effects on dominance at a point where the $_1A$ allele has maximal dosage and the dosage of the $_2A$ allele is very low.  Isoclines throughout represent intervals of 0.1.

**Figures:**

**Figure 1:**

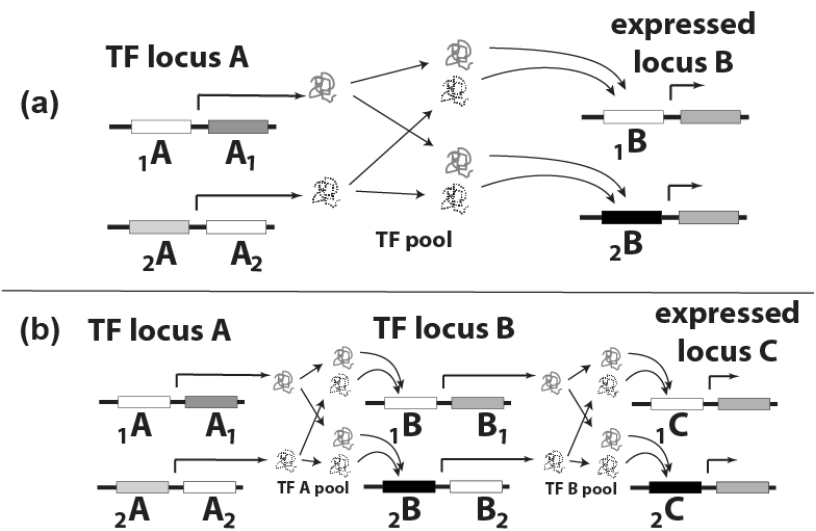

**Figure 2:**

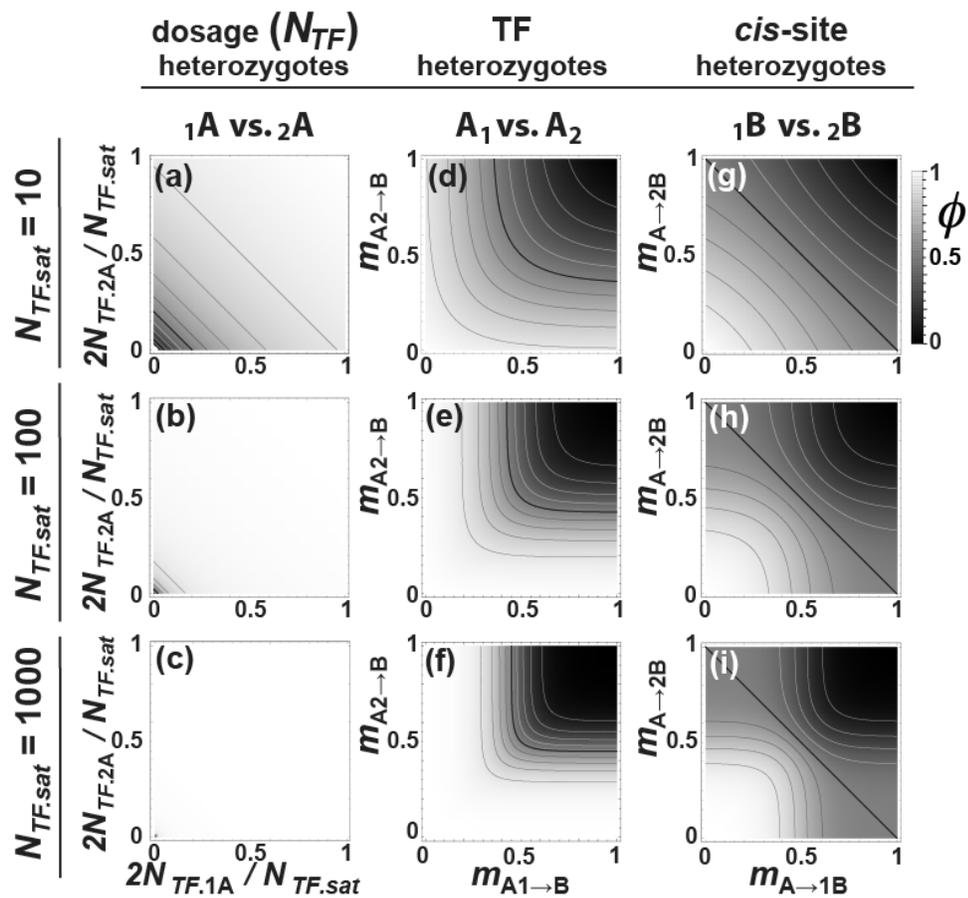

**Figure 3:**

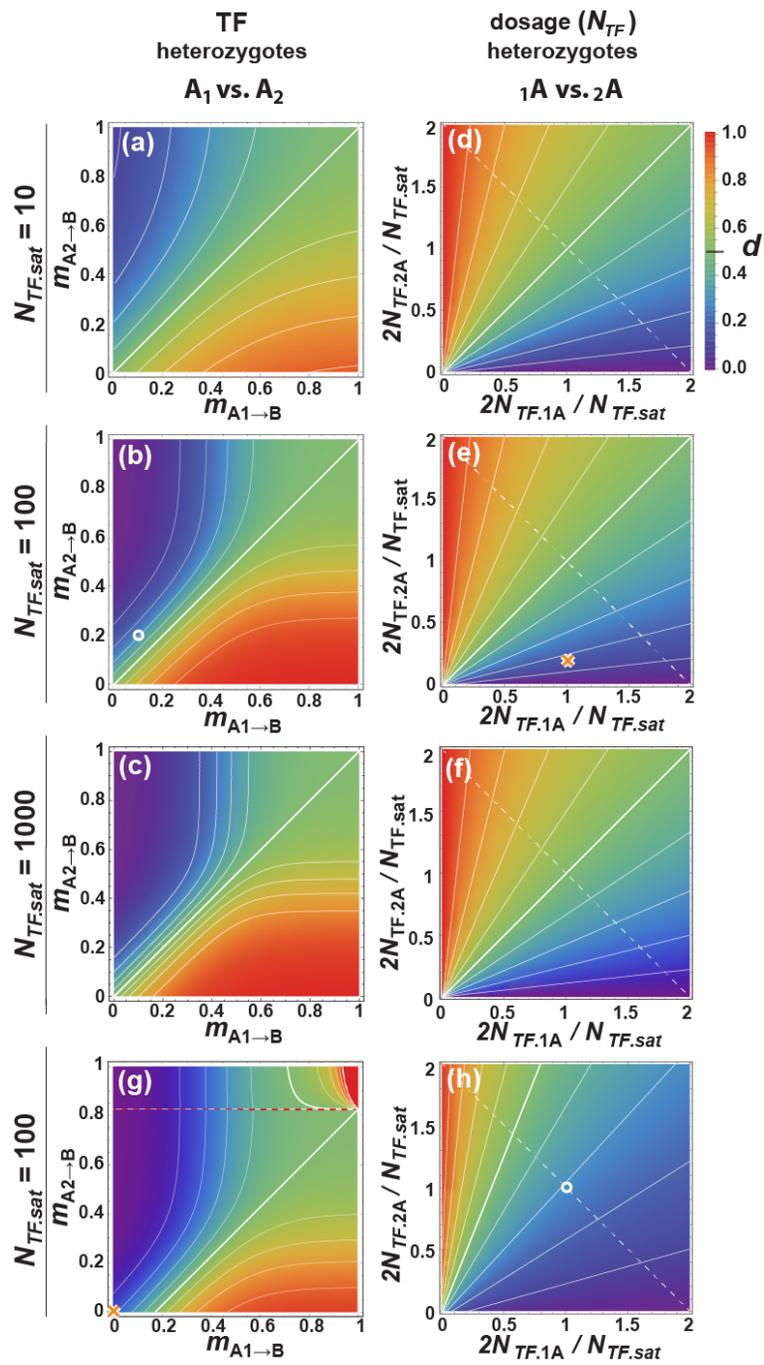

**Figure 4:**

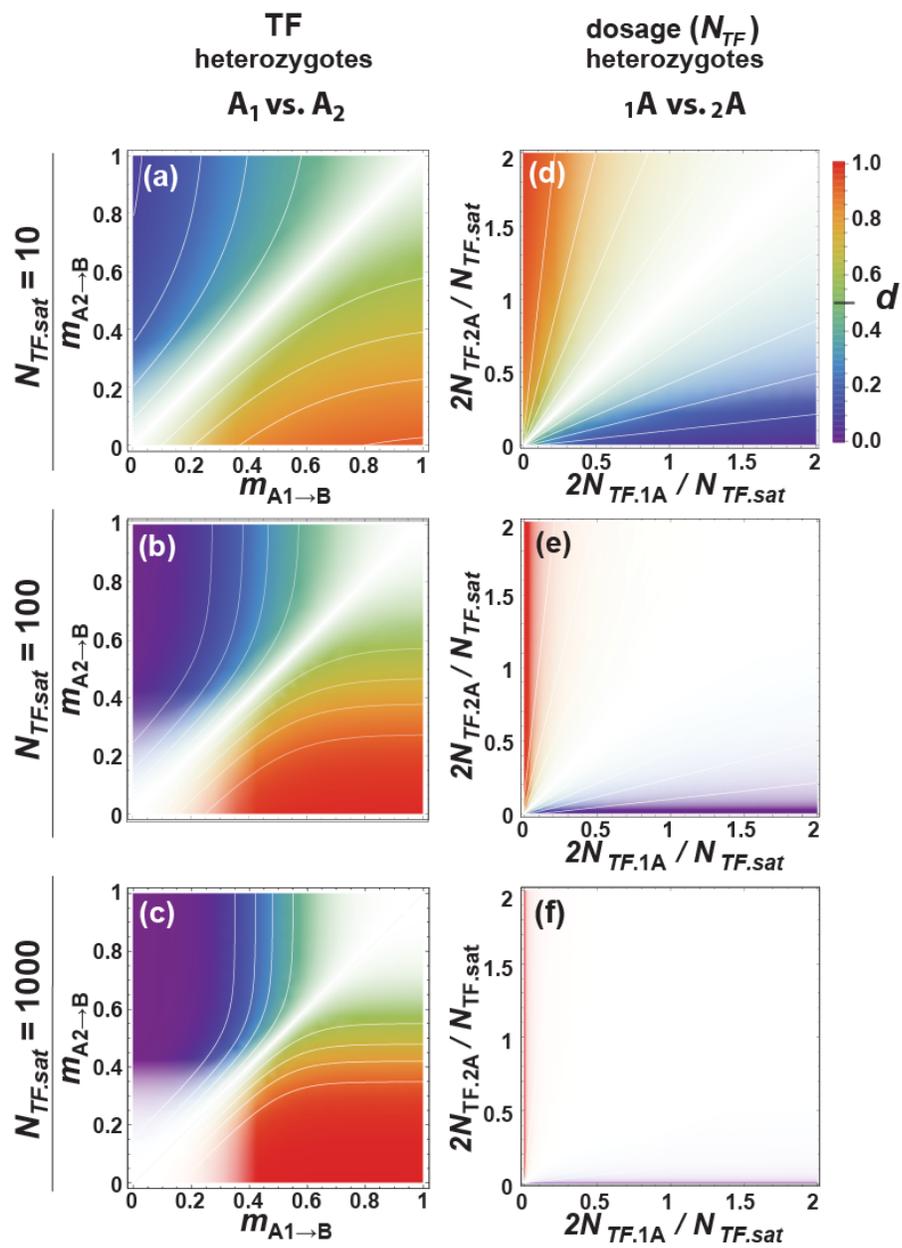

**Figure 5:**

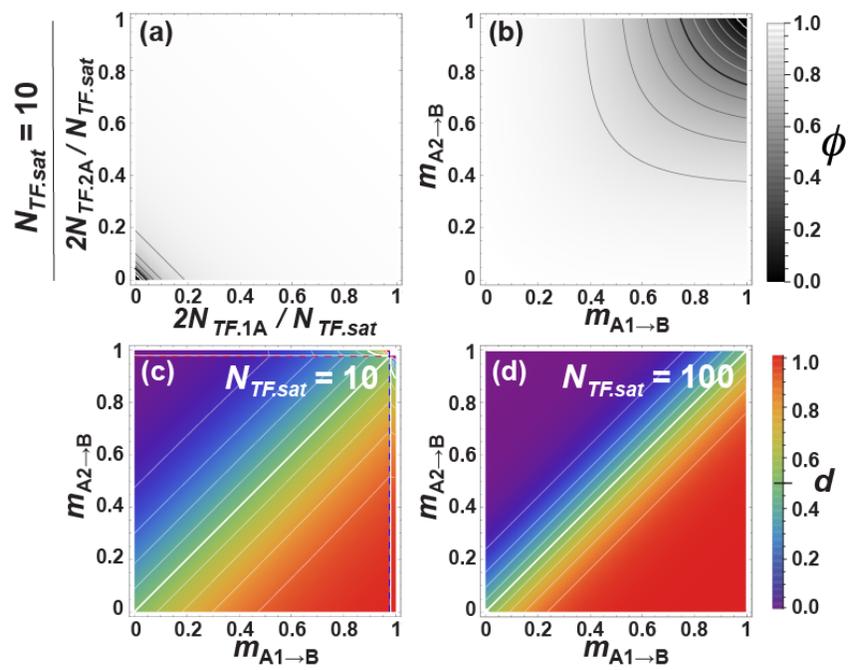

**Figure 6:**

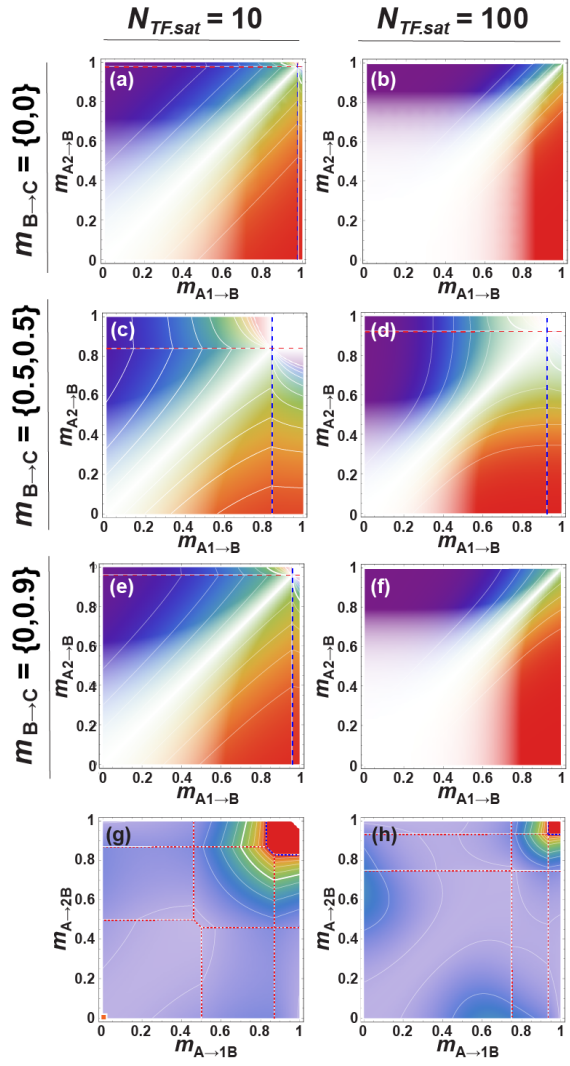

**Fig. 7**

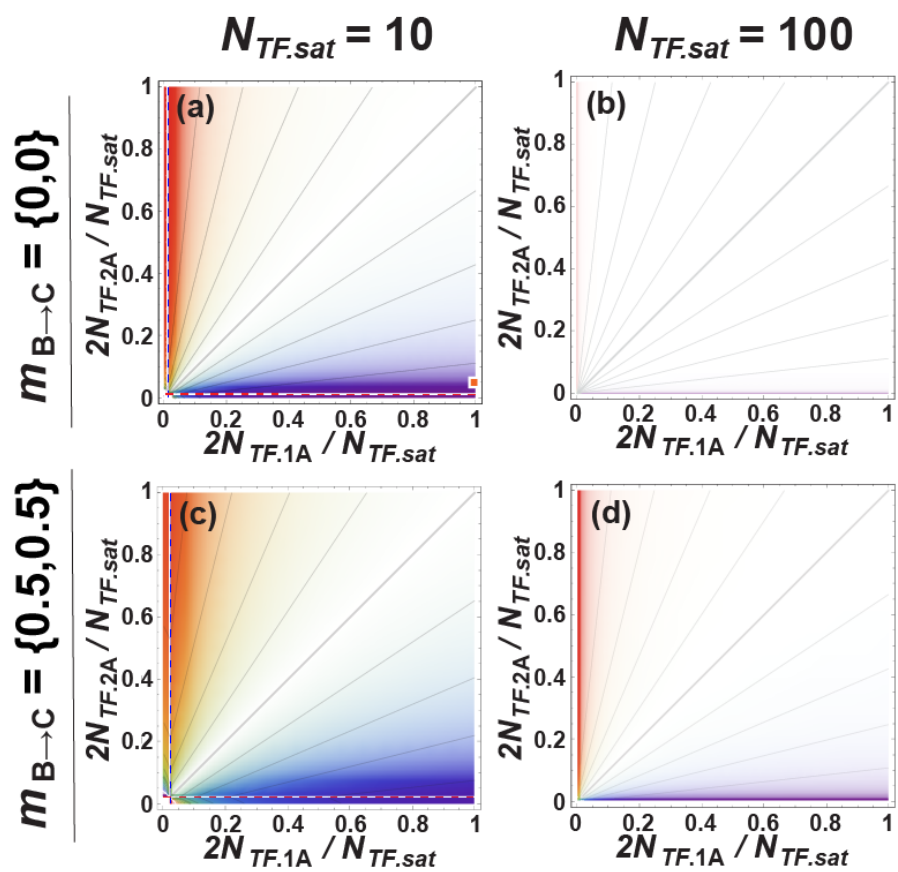

## Supporting Information

*Parameter reduction in the biophysical model*

The biophysical model we analyze is a parameter-reduced version of the model in Tulchinsky *et al.* (2014), developed from models of transcription-factor (TF) binding in the statistical physics literature (Gerland *et al.* 2002). The haploid version of that model characterizes fractional occupancy of the TF on the promoter site it regulates as

$$\theta = \frac{N'_{TF}}{N'_{TF} + \exp\left[-\Delta G + E_{diff}\right]}$$

where $\theta$ is the fractional occupancy, $N'_{TF}$ is the absolute number of TF molecules, $-\Delta G$ is the free energy of association between a TF molecule and promoter site, and $E_{diff}$ is the difference between the free energy of association between a TF molecule to its primary binding site and its local environmental background, which may include the non-specific binding to the genomic background as well as inhibitors and other molecules in the nuclear matrix (Mueller *et al.* 2013). When $E_{diff} < 0$, the background is more attractive and fewer TF's are available for gene regulation; when $E_{diff} > 0$, the target site is more attractive. Non-specific binding reduces the number of TF molecules in solution, making fewer available to interact with the specific binding site. We combine the $E_{diff}$ parameter and their $N'_{TF}$ into a single TF-availability term using $N_{TF} = N'_{TF}*\exp(-E_{diff})$, where $N_{TF}$ is the number of unencumbered TF molecules available for regulatory interactions, such that

$$\theta = \frac{N_{TF}}{N_{TF} + \exp\left[-\Delta G\right]}$$

Gerland *et al.* (2002) estimated that $E_{diff} = \sim 0$ or a little less, so in practice $N_{TF} = \sim N'_{TF}$ unless $N'_{TF}$ is very small.

The bioenergetic model represents the interacting TF molecules and promoter sequence as strings of bits, where binding decreases with $m'$, the number of mismatching bits. The second parameter modification we use is define a fractional mismatch parameter $m = m'/n$, where $n$ is the bitstring length. Therefore, our $-\Delta G_1$ is equivalent to $-n\Delta G_1$ of Tulchinsky *et al.* (2014). For resolution in our density plots, we treat $n$ as an arbitrarily large, finite integer. Reducing $n$ would increase pixilation in those plots by averaging over blocks of area $1/n^2$, without affecting the conclusions.

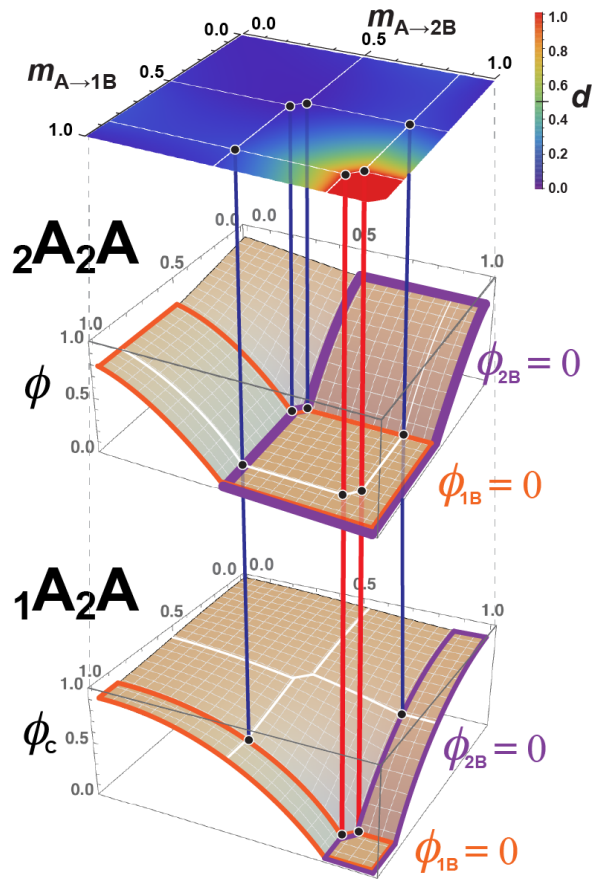

**Figure S1** Genetic-background effects on dominance in the $_1A_2A \rightarrow {}_1B_2B \rightarrow CC$ case: modification of the dominance of dosage allele $_1A$ due to binding variation in the *cis* site of TF locus B. The top image is the genotype-dominance map shown in Figure 6g under conditions $N_{TF.sat} = 10$, $N_{TF.1A} = N_{TF.sat}/2$ and $N_{TF.2A} = N_{TF.sat}/10$ (detectability mask omitted), i.e., where the dosage of the $_1A$ allele is maximal and that of the $_2A$ allele is very low. Axes in this map are the binding strengths $m_{A \rightarrow 1B}$ and $m_{A \rightarrow 2B}$, i.e., binding variation in the A→B step due to variation in the *cis* site of TF locus B, and they project to the images below it. Regions separated by dotted lines in Figure 6g are here outlined in white. Projected below this map are the associated 3-locus G-P maps ($\phi_C$'s) of the $_2A_2A$ homozygote and the $_1A_2A$ heterozygote with respect to variation at the *cis* site of TF locus B. For those genotypes, the purple lines bound the regions where $\phi_{1B} = 0$ (i.e., allele $_1B$ is not expressed in the A→B step, thus $N_{TF.1B} = 0$) and the orange lines bound the regions where $\phi_{2B} = 0$ (i.e., allele $_2B$ is not expressed in the A→B step, thus $N_{TF.2B} = 0$).